\documentclass[twocolumn,prl]{revtex4}

\usepackage{graphicx}

\begin{document}
\title{Dynamical heat channels}

\author{S. Denisov, J. Klafter, M. Urbakh}
\address{\it School of Chemistry, Tel-Aviv University, \it
Tel-Aviv 69978, Israel}

\date{\today}

\begin{abstract}
    We consider  heat conduction in a 1D dynamical channel. The
channel consists of a group of noninteracting particles, which
move between two heat baths according to some dynamical process.
We show that the essential thermodynamic properties of the heat
channel can be evaluated from the diffusion properties of the
underlying particles. Emphasis is put on the conduction under
anomalous diffusion conditions.
\end{abstract}
\maketitle

  The link between  thermodynamic phenomena
and microscopic dynamical chaos has been a subject of interest for
a long time [1]. Examples of such a relationship is deterministic
diffusion [2], where local dynamical properties, such as stability
(or instability) of several fixed points can change the global
diffusion transport from normal to anomalous [3]. Another question
which has attracted a lot of attention recently, is the problem of
heat conductivity in deterministic extended dynamical systems
[4-8]. A large number of models have been proposed in order to
understand the conditions under which  a system obeys the Fourier
heat conduction law [4]. Recently, a new class of 1D models,
"billiard gas channels", has been proposed [5-8]. These channels
consist of two parallel walls with a series of scatterers,
distributed along walls, and noninteracting  particles that move
inside. The two ends of the channel are  in contact with  the heat
baths. By changing the shapes and positions of scatterers it is
possible to change the conductivity of the channel [5-8].

In this Letter we show that such billiard gas channels belong to a
wider class of models, which we call {\it dynamical heat
channels}. The absence of interactions  between particles and the
independence of the particle dynamics on kinetic energy allow a
complete separation between the thermodynamic aspect, which is
governed by properties of the thermostats, and the dynamics inside
channel, which is governed by diffusion properties within the
channel. All the essential information on heat conductivity of
such a dynamical heat conductor can be obtained from the diffusion
properties of channel.

{\it Dynamical heat channel.}  To model dynamics within the
channel we consider  $N$ particles that move along direction $X$,
following the dynamical equations of motion
\begin{eqnarray} \label{1}
\dot{\mathbf{x}}=\mathbf{f}(\mathbf{x},t), ~~~~X \in \mathbf{x},
\end{eqnarray}
where the function $\mathbf{f}$ can be either deterministic  or
random.

To consider transport of heat, two heat baths with  temperatures
$T_{+}$ and $T_{-}$ are attached to the left and right ends of the
channel. The heat bath is characterized by a velocity probability
density function (pdf), $P_{T}(v_{Th})$, where $v_{Th}$ is the
thermal velocity. After colliding with the heat bath, the particle
is ejected back to the channel with a velocity $v_{Th}$ which is
chosen from $P_{T}(v_{Th})$.

As particles do not interact, the dynamics of the ensemble inside
the channel can be described by a long trajectory of a  single
particle and the flux should be rescaled by the factor $N$. The
trajectory of a particle is independent of the particle velocity
$v_{Th}$. The only difference between "hot" and "cold" particles
is that the hot ones cover the same trajectories faster then the
cold ones. The velocity $v_{Th}$ does not change during the
propagation through the channel and it can be interpreted as a
temperature "label" of a particle. The dependence of the dynamics
in Eq. (1) on $v_{Th}$
 can be taken into account by
introducing a scaling factor for the time  $t\rightarrow
t/v_{Th}$.

Due to the separation of the thermodynamic characteristics from
the dynamical ones, the proposed approach is not limited to
Hamiltonian systems only [5-8]. We only assume that the dynamics
inside the channel has a diffusional character, and can be
characterized by evolution of the  mean square displacement (msd)
\begin{eqnarray} \label{2}
\langle X^{2}(t)\rangle \sim t^{\alpha}.
\end{eqnarray}
This diffusion can be normal ($\alpha=1$), subdiffusive ($\alpha
<1$) or superdiffusive ($\alpha
>1$) [9].

Following  Ref. [5], heat transfer by a particle through the
channel  is
\begin{equation}
Q(t) = \sum_{j=1}^{M(t)} \Delta E_ j= \sum_{j=1}^{M(t)}q_{j} \cdot
(E_{j}^{in} - E_{j}^{out}),
\end{equation}
where $E_{j}^{in}$ and $E_{j}^{out}$ are the energies before and
after the $j$-collision with the heat bath. $q_{j}$ is the
direction factor, $q_{j}=1$ if the $(j-1)$-collision is with the
hot end, and $q_{j}=-1$ in the case of the cold end.  $M(t)$ is
the total number of collision events during time $t$. In the case
of normal heat conductivity, $Q$ grows linearly with $t$ and  the
heat flux is defined by
\begin{equation}
J =lim_{t\rightarrow \infty} \frac{Q(t)}{t}.
\end{equation}

Let us start from a situation where the particle is initially
located at the hot end. During diffusion it can come back and
collide with the hot bath again. According to Eq.(3) this event,
on average, does not lead to  heat transfer. But when the particle
reaches the opposite cold end $Q$ increases on average as
$\int_{0}^{\infty}
\frac{v_{Th}^{2}}{2}[P_{T_{+}}(v_{Th})-P_{T_{-}}(v_{Th})]dv_{Th}$.
After that the process is reiterated starting from the cold end.
Thus, the problem of heat transfer is reduced to the problem of
diffusion in a finite interval under reflecting and absorbing
boundary conditions. As the initial condition we assume that at
$t=0$ the particle is located at the reflecting end. The average
time $\tau$ needed to reach an absorbing boundary is the first
moment of the pdf $\phi(t)$ for first arrival times. To take into
account the effect of thermodynamic velocity on the  time $\tau$
should be rescaled, as mentioned above, $ \tau \rightarrow \tau
\int_{0}^{\infty} \frac{1}{v_{Th}} P_{T_{\pm}}(v_{Th})dv_{Th}$ .
Due to the absence of mass flux, the number of  transitions from
left to right and vise versa should be the same. Finally, we
obtain  the following equation for the one-particle heat flux
through the channel of length $L$
\begin{equation}
J(L) = \tau^{-1} \frac{\int_{0}^{\infty}
v_{Th}^{2}[P_{T_{+}}(v_{Th})-P_{T_{-}}(v_{Th})]dv_{Th}}{\int_{0}^{\infty}
\frac{1}{v_{Th}}[P_{T_{+}}(v_{Th})+P_{T_{-}}(v_{Th})]dv_{Th}}.
\end{equation}
For an ensemble of particles the heat flux should be written as
$J_{ens}=N \cdot J(L)$. Here we take into account that  in order
to keep a density of particles in the channel fixed,  $N$ should
be proportional to $L$; namely $N\propto L$.

Equation (5) demonstrates a complete separation between
thermodynamic and dynamical aspects. Without loss of generality we
consider  the "delta"-heat bath with a simple pdf,
$P_{T}(v_{Th})=\delta(v_{Th}-\sqrt{2T})$ [8].

As a model for the dynamics in the channel we consider continuous
time random walks (CTRW) [10] in a discrete lattice. This allows
to cover the spectrum of diffusion regimes, from subdiffusion to
superdiffusion, and to describe  kinetics of deterministic
Hamiltonian [11,12] and dissipative [3] systems.

The CTRW-model is  a stochastic process which represents an
alternating sequence of waiting and jumping events [10].  A
particle waits at each point for a time chosen from a waiting time
pdf $\psi_{w}(t)$, and makes a jump (flight) to the left or right
with equal probabilities. The jumps are characterized by the pdf
$\psi_{f}(x,t)$, the probability density to move  a distance $x$
at time $t$ in a a single fligth event. We consider power laws for
both pdf's[10]
\begin{equation}
\psi_{w}(t) \sim t^{-\gamma_{w}-1}, ~~~~ \psi_{f}(x,t) \sim
x^{-\gamma_{f}-1}.
\end{equation}
Depending on asymptotic properties of  $\psi_{w}(t)$ and
$\psi_{f}(x,t)$, we have regimes of superdiffusion, normal
diffusion or subdiffusion [10].

\begin{figure}[t]
\includegraphics[width=1.\linewidth,angle=0]{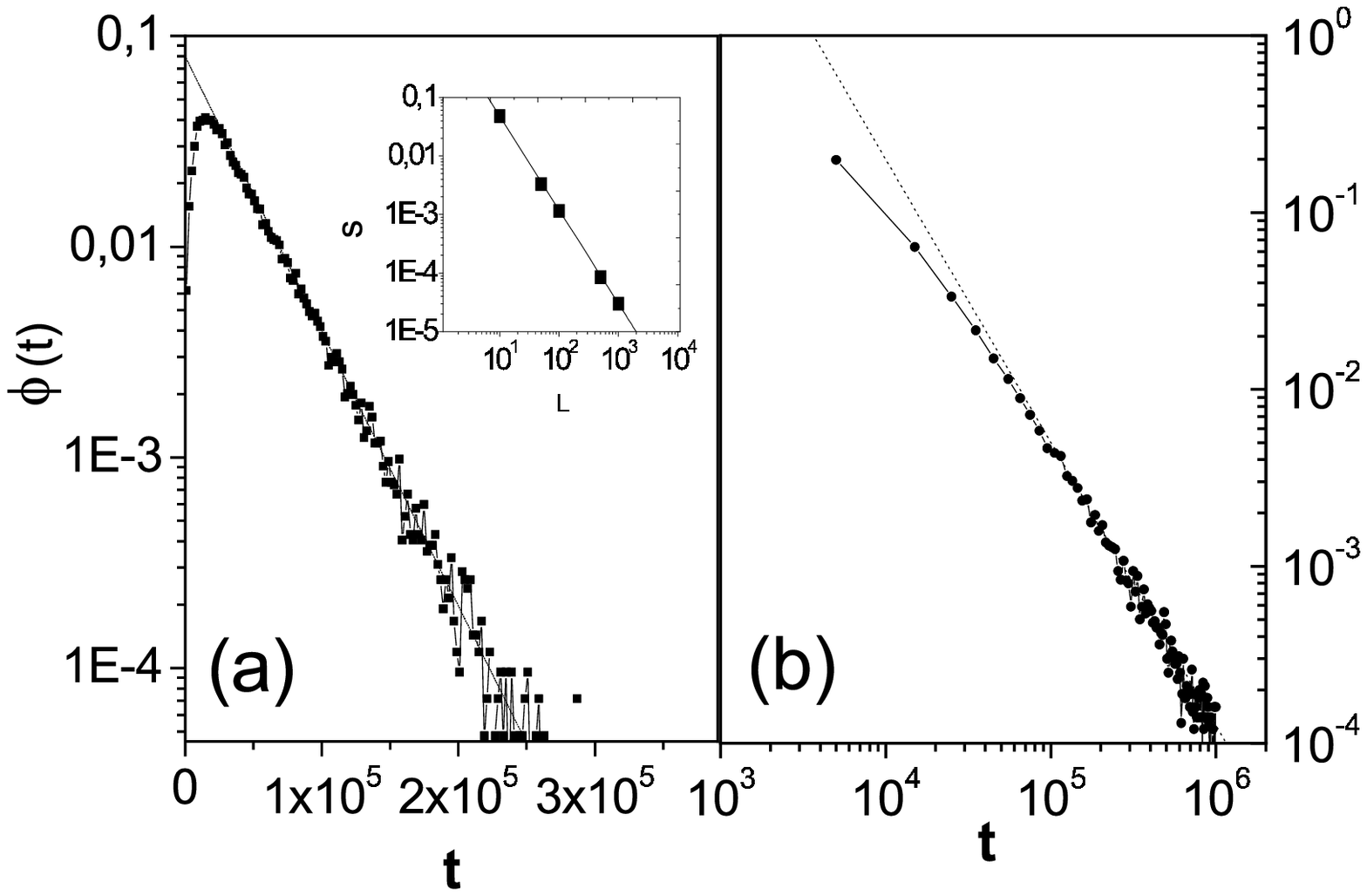}
\caption{ pdf $\phi(t)$ for  first arrival times for (a)
superdiffusion with $\gamma_{w}=2.6,~\gamma_{f}=1.6$ for $L=100$.
Inset showns the scaling of the exponent $s$ with the channel
length $L$, $L^{-\gamma_{f}}$.  (b) Subdiffusion,
$\gamma_{w}=0.6,~\gamma_{f}=2.6$. Straight lines correspond to the
asymptotic behaviour (see text). \label{fig2}}
\end{figure}

{\it Superdiffusion.} To achieve superdiffusion with a finite msd
$\langle x^{2}(t)\rangle$ there should be a correlation between
the length and duration of the individual flights. Such
correlation leads to the model of the L$\acute{e}$vy walks with
spatio-temporal pdf [10]
\begin{equation}
\psi_{f}(x,t)=\psi_{f}(t)\delta(|x|-vt),~~~~ \psi_{f}(t)\sim
t^{-\gamma_{f}-1}
\end{equation}
that corresponds to flights with a constant  velocity $v$. Here we
assume that $\gamma_{w}>1$, so there is a finite mean waiting
time. Depending on $\gamma_{f}$ we distinguish among three regimes
of diffusion  [10], according to the exponent $\alpha$ in Eq.(2),
\begin{equation}
  \alpha =
  \left\{\begin{array}{ll} 2, ~~~~~~~~0<\gamma_{f}<1 \\
      3-\gamma_{f},~~1<\gamma_{f}<2, \\
      1, ~~~~~~~~~ 2<\gamma_{f}
    \end{array}\right.
\end{equation}

In order to derive the dependence of the heat flux $J$ on the
length $L$ of the channel  we start from a consideration of the
survival probability $\Phi (t)$ for a particle walking on the
finite interval of the length $L$ bounded by reflecting and
absorbing boundaries. We find that for L$\acute{e}$vy walks the
survival probability is $\Phi(t) \propto e^{-st}$, where $s
\propto L^{-1}$ for the ballistic motion ($\alpha=2$), $s \propto
L^{-\gamma_{f}}$ for superdiffusion ($1< \alpha < 2$) , and $s
\propto L^{-2}$ for normal diffusion ($\alpha=1$)(see inset in
Fig1a). Then, the pdf for first arrival times, $\phi(t)$, is
$\phi(t) \sim -\dot{\Phi}(t) \propto e^{-st}$ (Fig.1a) and has a
finite first moment, $\tau$. As a result, in asymptotic limit the
heat, $Q$, grows linearly with time $t$, $Q(t) \sim (\Delta
E/\tau)t$ (line (3) in Fig.2), where the mean arrival time is
$\tau=\int_{o}^{\infty}t \phi(t)dt  \propto s^{-1}$ (Fig.3a).
Thus, in the case of superdiffusion, due to L$\acute{e}$vy walks,
we get a normal linear in time heat conductivity.

\begin{figure}[t]
\includegraphics[width=0.8\linewidth,angle=0]{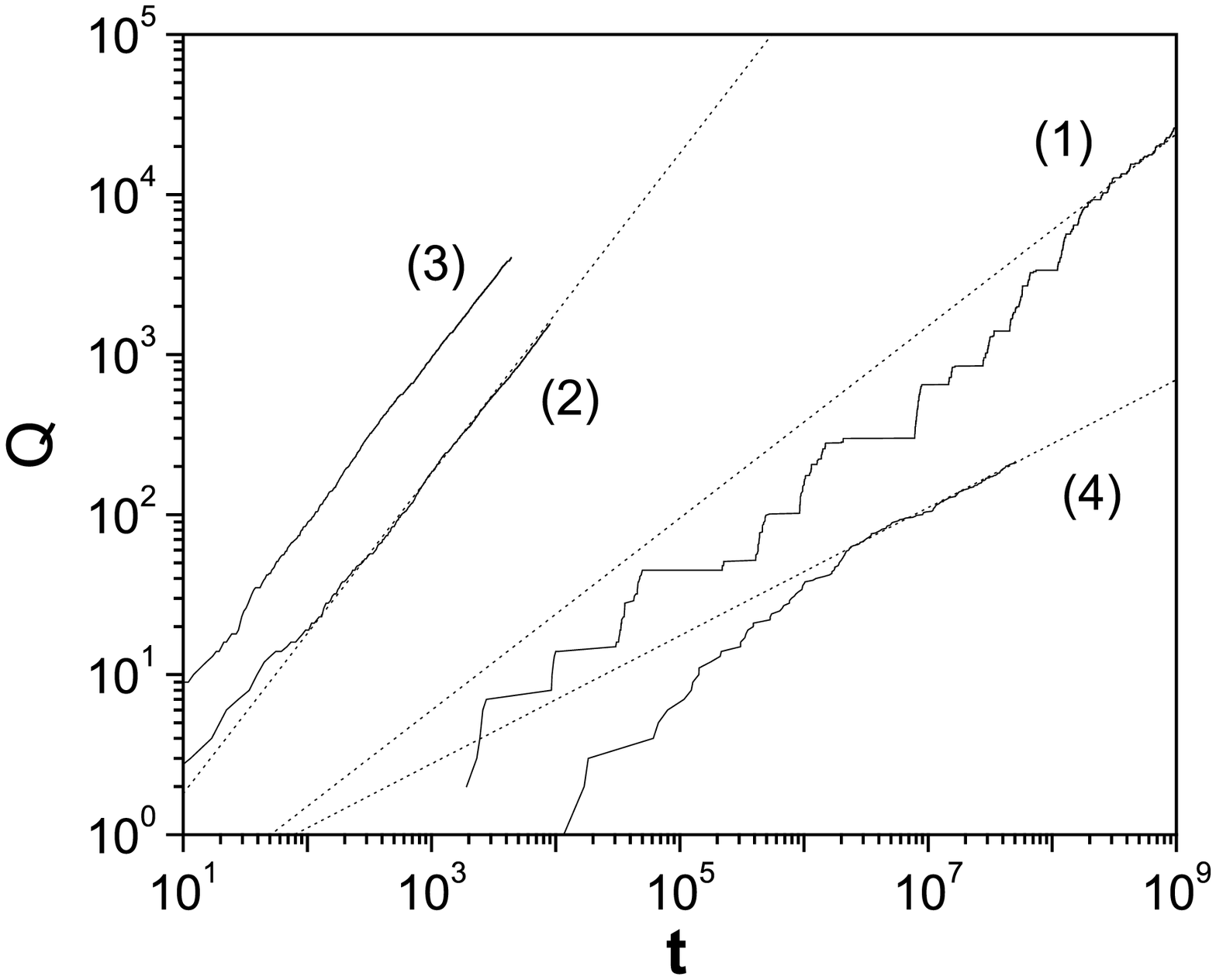}
\caption{ Evolution $Q(t)$ vs $t$ for subdiffusion,
$\gamma_{w}=0.6,~\gamma_{f}=2.6$ (curve (1)), normal diffusion,
$\gamma_{w}=1.6,~\gamma_{f}=2.4$ (curve (2)), and superdiffusion,
$\gamma_{w}=1.6,~\gamma_{f}=1.6$ (curve (3)). Curve (4)
corresponds to rescaled, for one particle, $Q(t)$ for an ensemble
$N=100$ in the case of competition between anomalous flights and
localized events, $\gamma_{w}=0.4,~\gamma_{f}=1.2$. Straight lines
correspond to asymptotics (see text)  For all cases the length of
channel is $L=1000$, $T_{+}=2$ and $T_{-}=1$. \label{fig2}}.
\end{figure}

 Finally, for the one-particle heat flux, following Eqs. (5) and (8),we arrive at
the following equation for $J(L)$ in term of the msd exponent
$\alpha$,
\begin{equation}
  J(L)\propto L^{-\beta},~~~ \beta =
  \left\{\begin{array}{ll} 1, ~~~~~~~~~~~~~~~ \alpha=2 \\
      \gamma_{f}= 3-\alpha,~~1<\alpha<2, \\
      2, ~~~~~~~~~~~~~~~ \alpha=1
    \end{array}\right.
\end{equation}
In Fig3.b we show the numerical results for $J(L)$ obtained for
$\gamma_{f}=1.6$ (stars) and $\gamma_{f}=2.4$ (squares).  The
results are in good agreement with the scaling law suggested by
Eq.(9) (straight lines in Fig.3b).

Then taking Eq.(9) into consideration the thermal conductivity
$k=-J_{ens}(N)/\nabla T$,
 shows the following behavior for
$1<\alpha<2$,
\begin{equation}
k \propto L^{2-\gamma_{f}} \propto L^{\alpha - 1}.
\end{equation}
This diverges as one goes to the thermodynamic limit $L\rightarrow
\infty$. For the ballistic case $\alpha=2$, $k\propto L$ and for
$\alpha=1$ $k\propto const$ as expected.

The L$\acute{e}$vy walk model is an adequate approach for modeling
of Hamiltonian kinetic in a mixed phase space [12]. Previous
numerical results which have been obtained for Hamiltonian
billiard channels, show that $\alpha=1.3$ corresponds to the flux
exponent $\beta=1.72$ [7], and $\alpha=1.8$ to $\beta=1.178$ [8],
which are in good agreement with the relation in Eq.(9).

In the case of L$\acute{e}$vy flights [10], unlike the
L$\acute{e}$vy walks, the velocity is not introduced explicitly.
Here we assume that all flights have the same duration $t_{f}$ and
use the following decoupled representation for flights pdf,
\begin{equation}
\psi_{f}(x,t)=\psi_{f}(x)\delta(t-t_{f}),~~~~ \psi_{f}(x) \sim
x^{-\gamma_{f}-1}.
\end{equation}

\begin{figure}[t]
\includegraphics[width=1.\linewidth,angle=0]{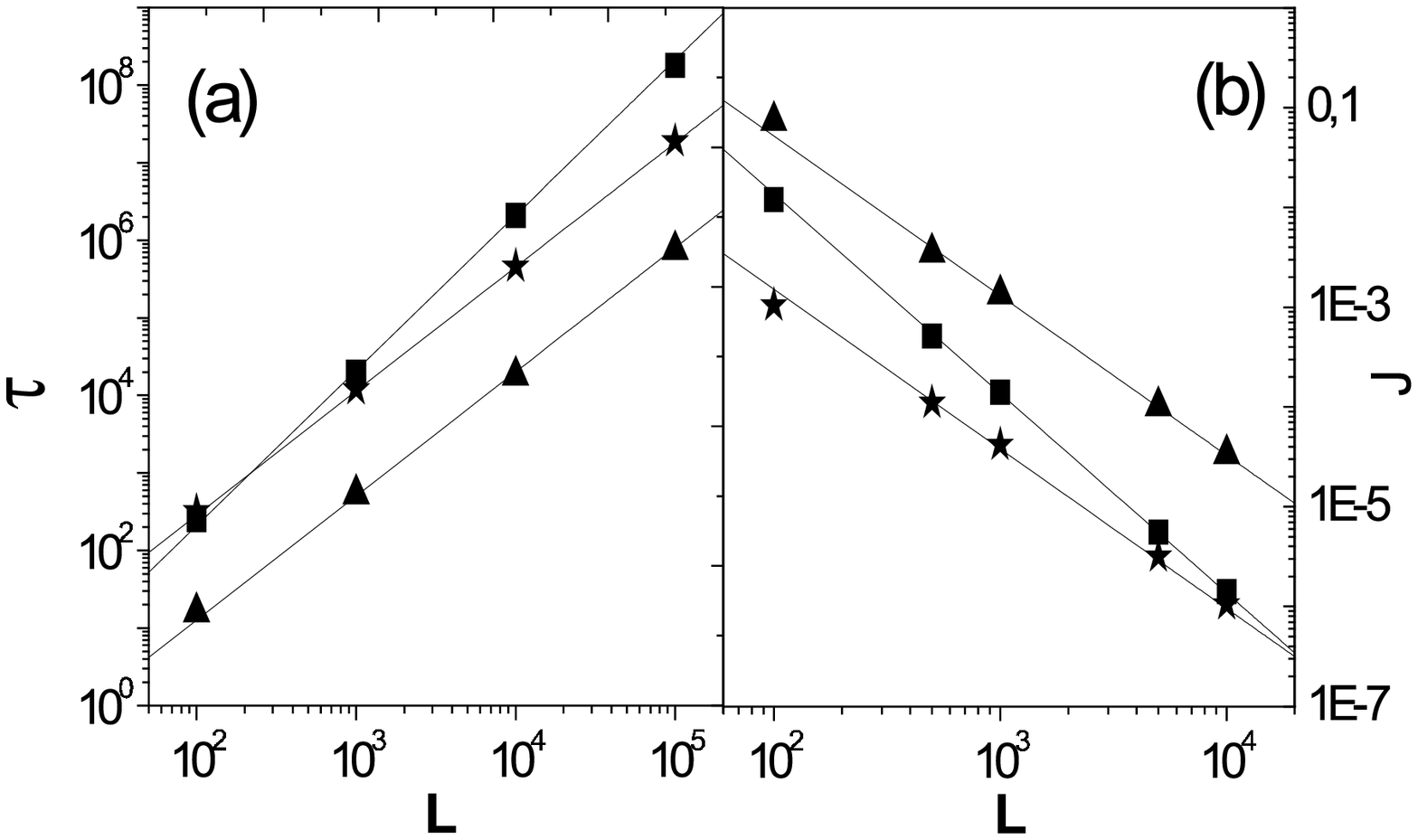}
\caption{ (a) The mean first arrival time $\tau$ versus $L$; (b)
one-particle flux $J(L)$, Eq. (5), versus $L$ ( $T_{+}=2$ and
$T_{-}=1$). Results correspond to different diffusion regimes with
subdiffusion, $\gamma_{w}=0.6,~\gamma_{f}=2.6$ (circles), normal
diffusion $\gamma_{w}=1.6,~\gamma_{f}=2.4$ (squares),
superdiffusion $\gamma_{w}=1.6,~\gamma_{f}=1.6$ (stars), and
L$\acute{e}$vy flights, $\gamma_{w}=1.6,~\gamma_{f}=1.6$
(triangles). Straight lines correspond to asymptotics (see text).
 \label{fig1}}
\end{figure}

In this case the msd diverges and one is tempted to use $\langle
|x(t)| \rangle^{2}$ to characterize the dynamics. In the
asymptotic regime this quantity scales as $\langle |x(t)|
\rangle^{2} \sim t^{2/\gamma_{f}}$[10]. As in the previous case,
here the exponent $\gamma_f$  determines the dynamics in the
channel and the dependence of flux $J(L)$ on $L$.  The scaling of
the mean arrival time, $\tau \propto L^{\gamma_{f}}$, is
consistent with the result obtained from a direct solution of a
fractional Fokker-Planck equation [14]. We note that  using of
$\langle |x(t)| \rangle^{2}$ instead of msd gives a different, yet
unphysical scaling, $ J(L) \propto L^{ \frac{2}{\gamma_{f}}-3}$,
with an exponent which is close to that in the scaling in Eq. (9)
and coincides with it at the points $\gamma_{f}=2$ (normal
diffusion) and $\gamma_{f}=2$ (ballistic diffusion).

{\it Subdiffusion.} In the case of subdiffusion $\gamma_{w}<1$ and
$\gamma_{f}>2$ in Eq.(6) and the mean waiting time $\tau_{w}$
diverges [10]. The motion is characterized by long localized
events and the msd grows sublinearly, $\langle x^{2}(t)\rangle
\sim t^{\gamma_{w}}$. The pdf of the first arrival time,
$\phi(t)$, has a power-law asymptotics $\phi(t)\propto
t^{-\gamma_{w}-1}$ (Fig.1b). This finding is in line with a result
for subdiffusion under a bias in a seminfinite interval with an
absorbing boundary [15]. The anomalous character of the pdf
$\phi(t)$ leads to a divergence of $\tau$. Thus, in the case of
subdiffusion the heat $Q(t)$ carried through the channel grows
sublinearly with time (see line (1) in Fig.2),
\begin{equation}
Q(t)\sim t^{\gamma_{w}}.
\end{equation}
Within a traditional definition of the flux, Eq.(4), the
subdiffusive anomalous heat conductivity can not be distinguished
from a heat isolator.
 We note that subdiffusion regimes can not be achieved in the case
of channels with Hamiltonian dynamics due to a finiteness of
recurrence time [9].

\begin{figure}[t]
\includegraphics[width=0.9\linewidth,angle=0]{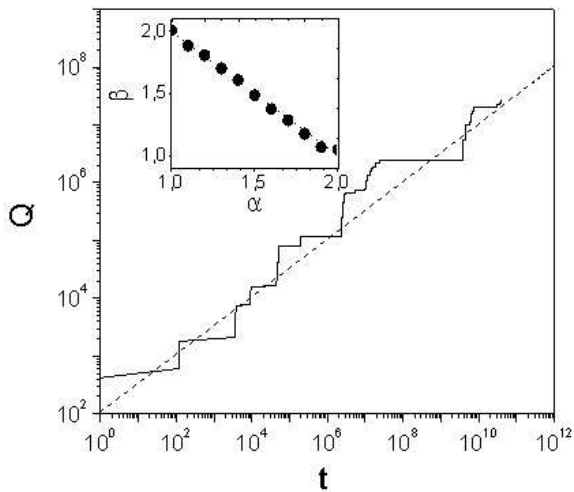}
\caption{Evolution $Q(t)$ vs $t$ for the map in Eq. (14) with
$\gamma_{w}=0.5,~\gamma_{f}=1.5$ ($L=1000$, $dt=10^{-4}$,
$T_{+}=2$ and $T_{-}=1$). Straight lines correspond to power law
dependence in Eq. (12). Inset shown numerically obtained relation
between scaling exponent $\beta$ for one-particle flux and
exponent $\alpha$ ($\gamma_{w}=1.6$ for all cases). Straight lines
correspond to $\beta=3-\alpha$. \label{fig2}}
\end{figure}

{\it Coexistence of long localized events and fligths.}  In the
case of  competition between fligths and localization events the
msd exponent $\alpha$ is given by a relation [10]
\begin{equation}
\alpha=2+\gamma_{w}-\gamma_{f},~~~0<\gamma_{w}<1,~ 1<\gamma_{f}<2.
\end{equation}

The pdf for the arrival time, $\phi (t)$, is governed, like in the
case of simple subdiffusion, by the pdf for waiting times,
$\psi_{w}(t)$. The divergence  of mean arrival time $\tau$ can
lead to anomalous heat conductivity, Eq. (12), even in the case of
superdiffusion spreading, $\langle x^{2}(t)\rangle \sim
t^{\alpha}$, $\alpha > 1$, Eq. (13). Fig.1 (line (4)) shows the
dependence $Q(t)$, for the ensemble of $N=100$ particles
($\gamma_{w}=0.4, \gamma_{f}=1.2$, $\alpha=1.2$), rescaled for one
particle.

One can go continuously among the various diffusional behaviors by
tunning parameters in an iterated map. As an example of
deterministic channel we consider a combined map [3]. This
one-dimensional map generates intermittent chaotic motion with
coexisting localized and ballistic motion events. The map is
defined for one unit cell by the iterative rules,
\begin{eqnarray}
X_{n+1}=f(X_{n}),~~~~~~~~~~~~~~~~~~\nonumber \\
  f(X)=
  \left\{\begin{array}{ll}X+aX^{z}-1, ~~~~~0<X<\frac{1}{4}, \\
      X-\tilde{a}(\frac{1}{2}-X)^{\tilde{z}},
      ~~~\frac{1}{4}<X<\frac{1}{2}.
   \end{array} \right.
\end{eqnarray}

The parameters are $a=4^{z}\tilde{z}/(z+\tilde{z})$ and
$\tilde{a}=4^{\tilde{z}}z/(z+\tilde{z})$. The localized and
ballistic phases are characterized by the pdf's in Eq. (6),  with
the exponents $\gamma_{f}=(z-1)^{-1}$ and
$\gamma_{w}=(\tilde{z}-1)^{-1}$. Variation of the map parameters
$z$ and $\tilde{z}$ allows to cover the diffusional regime in the
range $0<\alpha \leq 2$ .

The thermodynamic velocity, $v_{Th}$, can be included into the map
description. We fix a time step $dt$ and after one iteration of
the map, Eq.(14), stretch the system time following
$t=t+dt/v_{Th}$. The length of channel is determined by a number
of unit cells, $0>[X]>L$. When the particle reaches the boundary
cells, $[X]=0$ or $[X]=L$, it is randomly placed into the interval
$0< X <1/2$ or $L-1/2< X <L)$ respectively. This corresponds to a
diffusive reflection of particle back to the channel.

The values of the flux exponents $\beta$  obtained for the map in
Eq. (14) are in agreement with the prediction in Eq.(9) for
superdiffusion (see inset in Fig.4). In Fig4. we show the time
evolution of $Q(t)$ for the case of  subdiffusion,
$\gamma_{w}=0.5$, $\gamma_{f}=1.5$. The steps in the $Q(t)$
dependence correspond to anomalously long waiting events when
particle is trapped near marginal stable fixed points
$x=j+\frac{1}{2}, j=0,...., L$ [3].

In summary we have introduced a class of dynamical heat
conductors. This class includes as  particular cases  recently
proposed Hamiltonian billiard channels [5-8]. In the absence of
interactions between the particles and the independence of the
dynamics on particle energy, the proposed aproach goes beyond
Hamiltonian dynamics and allows to express heat conductivity in
terms of channel diffusion properties. The Hamiltonian character
of dynamics becomes essential when interactions between particles
are introduced (like in the case of dynamical lattices [4]) or a
dependence of dynamics on particle energy is included.

Financial support from the Israel Science Foundation, the US
Israel BSF, and INTAS grants is gratefully acknowledged.


\begin{thebibliography}{99}
\bibitem {13} For generation of random variable $x$ with distribution in Eq.(7) we have used the
random variable $\xi$ with uniform distribution in unit interval
and transformation $x =t_{c}\xi^{-\frac{1}{\gamma}}$.
\end{thebibliography}
\end{document}